\documentclass[showpacs,prl,twocolumn]{revtex4}%
\usepackage{amssymb}
\usepackage{amsmath}
\usepackage{graphicx}
\usepackage{amsfonts}
\begin{document}
\title{Unified description of bulk and interface-enhanced spin pumping}
\author{S. M. Watts}
\author{J. Grollier}
\altaffiliation[Now at ]{Unit\'{e} Mixte de Physique CNRS/Thales, Route d\'{e}partementale 128, 91767
Palaiseau Cedex, France.}
\author{C. H. van der Wal}
\author{B. J. van Wees}
\affiliation{Physics of Nanodevices, Materials Science Center, University of Groningen,
Nijenborgh 4, 9747 AG Groningen, The Netherlands}
\date{\today }

\begin{abstract}
The dynamics of non-equilibrium spin accumulation generated in metals or
semiconductors by rf magnetic field pumping is treated within a diffusive
picture. \ The dc spin accumulation produced in a uniform system by a rotating
applied magnetic field or by a precessing magnetization of a weak ferromagnet
is in general given by a (small) fraction of $\hbar\omega$, where $\omega$ is
the rotation or precession frequency. \ With the addition of a neighboring,
field-free region and allowing for the diffusion of spins, the spin
accumulation is dramatically enhanced at the interface, saturating at the
universal value $\hbar\omega$ in the limit of long spin relaxation time.
\ This effect can be maximized when the system dimensions are of the order of
$\sqrt{2\pi D/\omega}$, where $D$ is the diffusion constant. \ We compare our
results to the interface spin pumping theory of A. Brataas \textit{et al}.
[Phys. Rev. B \textbf{66}, 060404(R) (2002)].

\end{abstract}

\pacs{85.75.-d,72.25.Mk,75.70.Cn}
\maketitle

A new direction in the field of spin electronics is the study of the
interaction between spin currents and the magnetization dynamics of a
ferromagnetic electrode into, or out of which the spin currents flow. \ In
spin pumping devices \cite{berger,brataas,sharma}, electron spins are pumped
from a ferromagnet in resonance with an rf magnetic field, into an adjoining
non-magnetic metal without accompanying charge currents. The so-called spin
battery \cite{brataas} is based on the interface spin-mixing conductance
\cite{mixcond}, which describes the coherent reflection and spin rotation
within an exchange length (typically a few nm in conventional ferromagnets) of
the interface. \ It has been experimentally verified by measuring the effect
of adjoining metal layers on the Gilbert damping constant of a ferromagnetic
layer \cite{damping,mizukami}, due to the transfer of angular momentum away
from the ferromagnetic and into adjacent metals \cite{tserkovnyak}. \ For the
spin battery, under specific conditions the universal value of $\hbar\omega$
is obtained for the spin accumulation, where $\omega$ is the precession
frequency of the magnetization. \ On the other hand, a bulk-oriented approach
based on mean-field theory \cite{tserkovnyak2} has been used to relate the
intrinsic Gilbert damping to the spin relaxation processes resulting from the
generation of non-equilibrium spin accumulation in the ferromagnet.

We present a model in which spin pumping with a rotating magnetic field can
produce spin accumulation in bulk material. \ Moreover, in a hybrid system in
which spins are allowed to diffuse from a "pumped" region into a neighboring
region without fields, the spin accumulation can be dramatically enhanced near
the interface. \ In both cases the spin accumulation reaches the universal
limit $\hbar\omega$ for long spin relaxation times. Our approach is based on
classical dynamics of spin ensembles in a diffusive system, and is applicable
to spin pumping in metals and semiconductors driven by a rotating, applied
magnetic field, as well as in weak ferromagnets driven by magnetization
precession. \ However, we emphasize that our semiclassical approach cannot be
applied directly to the conventional strong ferromagnet regime in which the
large exchange fields require a non-diffusive, quantum mechanical treatment
\cite{mixcond} because of the very small magnetic coherence length, much less
than the mean-free-path. \ The dc spin-accumulation induced by spin-pumping is
first derived\ for a bulk, uniform system, for which we obtain the surprising
result that bulk spin accumulation is generated even for small Zeeman
splitting. \ Then, by connecting a magnetic field-free region and solving the
boundary-value problem of diffusion across the interface, we find a transition
from bulk to interface-enhanced spin accumulation.

We start with Bloch-type equations written in terms of the chemical potentials
for the three spin directions, $\vec{f}=(f_{x},f_{y},f_{z})$ that describe the
space- and time-dependent, non-equilibrium spin accumulation $\vec{f}(x,t)$
(restricting the description to one spatial dimension $x$)
\cite{magnetization}. \ The basis of our description is the following equation
for the dynamics of $\vec{f}(x,t)$ in a rotating magnetic field $\vec{B}(x,t)$
\cite{johnson88,hernando}:
\begin{equation}
-\frac{\partial\vec{f}}{\partial t}+\vec{I}(x,t)=-D\nabla^{2}\vec{f}%
+\frac{\vec{f}}{\tau}+\frac{g\mu_{B}}{\hbar}\vec{B}\times\vec{f}%
,\label{Bloch1}%
\end{equation}
where $D$ is the diffusion constant and $\vec{I}(x,t)$ is a source term which
will be described in the next paragraph. \ The length scales relevant to the
problem are the spin diffusion length $\lambda=\sqrt{D\tau}$, the Larmor
precession length $\lambda_{B}=\sqrt{2\pi D/|\vec{\omega}_{B}|}$, where
$\hbar\vec{\omega}_{B}=g\mu_{B}\vec{B}$, and a length $\lambda_{\omega}%
=\sqrt{2\pi D/\omega}$ corresponding to the distance the spins can diffuse in
one period $\frac{2\pi}{\omega}$. \ Eqn. \ref{Bloch1}, with the left hand side
equal to zero, has been shown to give an accurate description of spin dynamics
in metallic systems for time-independent magnetic fields
\cite{johnson88,jedema2}.

The key new ingredient of our approach is the source term:%

\begin{equation}
\vec{I}(x,t)=-g\mu_{B}\frac{\partial}{\partial t}\vec{B}(x,t).\label{Is}%
\end{equation}
It describes the rate at which the oscillating magnetic field, via the Zeeman
energy, pushes spins aligned with the magnetic field below the Fermi level,
and anti-aligned spins above from the Fermi level. \ This mechanism can be
seen as a source of locally-injected, time-dependent spin currents
\cite{quantum}. \ Without this source term, $\vec{f}=0$ in the steady state.

The role of Eqn. \ref{Is} is sketched out as follows: \ consider two magnetic
fields applied in orthogonal directions, say $B_{x}$ and $B_{y}$, equal in
magnitude but oscillating out of phase by $\pi/2$: $B_{x}=B_{xy}\cos\omega t$
and $B_{y}=B_{xy}\sin\omega t$. \ The resulting $I_{x}$ generates spin
accumulation in $f_{x}$ in phase with $B_{y}$, and subsequent precession
yields spin accumulation in $f_{z}$. Likewise, $I_{y}$ produces an $f_{y}$
that is $\pi$ out of phase with $B_{x}$, and subsequent precession yields spin
accumulation in $f_{z}$ with the same sign as for the $I_{x}$ case. Both
$f_{x}$ and $f_{y}$ oscillate about a mean of zero with frequency $\omega$.
However, $f_{z}$ acquires a finite dc component. \ The steady-state solution
of $\vec{f}$ can be imagined as a spin accumulation vector that follows the
rotating field vector, lagging by phase angles which depend on $\omega$ and
$\tau$.
\begin{figure}[ptb]
\begin{center}
\includegraphics[height=3.1038cm,width=7.3455cm]{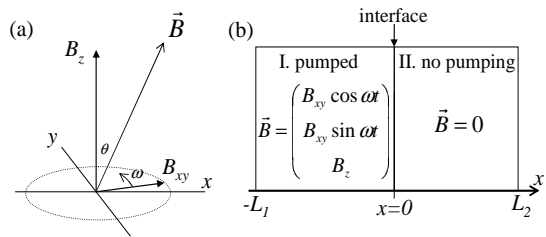}
\caption{(a)\ Magnetic field vector diagram. \ An applied rf field $B_{xy}$
rotates with angular frequency $\omega$ in the x-y plane. \ A static magnetic
field $B_{z}$ is applied perpendicular to $B_{xy}$ generating a field vector
$\vec{B}$ that rotates around $B_{z}$ with cone angle $\theta$. \ (b) The
general hybrid system with an interface at $x=0$. \ The fields just described
are applied in the "pumped" region I, $x<0$. \ $\vec{B}=0$ in region II,
$x>0$.}%
\label{diagram}%
\end{center}
\end{figure}

Fig. \ref{diagram}a shows the basic field configuration we will consider: \ a
rotating magnetic field $B_{xy}$ and a perpendicular, static field $B_{z}$.
\ The field vector is $\vec{B}=(B_{xy}\cos\omega t,B_{xy}\sin\omega t,B_{z})$
in Cartesian coordinates. \ We first consider region I (Fig. \ref{diagram}b)
alone; \textit{i.e.}, a uniform (spatially invariant) system without
boundaries. Dropping the diffusion term and rewriting Eqn.\ref{Bloch1} in a
more concise form:
\begin{equation}
\tau\dot{\vec{f}}=-\vec{f}-\vec{\omega}_{B}\tau\times\vec{f}-\hbar\dot
{\vec{\omega}}_{B}\tau,\label{Bloch2}%
\end{equation}
where $\vec{\omega}_{B}=(\omega_{xy}\cos\omega t,\omega_{xy}\sin\omega
t,\omega_{z})$. \ Since the induced spin accumulations $f_{x}$ and $f_{y}$
will, in the steady state, oscillate with the same frequency as $B_{x}$ and
$B_{y}$, it will be advantageous to transform to a coordinate system in which
the $x$ and $y$ axes rotate about the $z$ axis with angular frequency
$\vec{\omega}=\omega\hat{z}$, so that $\vec{B}=(B_{xy},0,B_{z})$. The time
derivative of a vector $\vec{r}$ in the stationary frame transforms as
$(\dot{\vec{r}})_{lab}\Rightarrow(\dot{\vec{r}})_{rot}+\vec{\omega}\times
(\vec{r})_{rot}$. Using this relation, Eqn.\ref{Bloch2} now becomes
\begin{equation}
\tau\dot{\vec{f}}=-\vec{f}-(\vec{\omega}_{B}+\vec{\omega})\tau\times\vec
{f}-\hbar(\dot{\vec{\omega}}_{B}+\vec{\omega}\times\vec{\omega}_{B}%
)\tau,\label{BlochRot}%
\end{equation}
In the last term, we have $\dot{\vec{\omega}}_{B}=0$ in this frame and
$\vec{\omega}\times\vec{\omega}_{B}=(0,\omega\omega_{xy},0)$. \ For the
stationary analysis we set $\dot{\vec{f}}=0$, and obtain the result
\cite{magnetization}:%
\begin{equation}
f_{z}=\frac{(\omega_{xy}\tau)^{2}}{1+(\omega_{xy}{}\tau)^{2}+((\omega
_{z}+\omega){}\tau)^{2}}\hbar\omega.\label{fzU}%
\end{equation}
The other components are related to $f_{z}$: $\ f_{x}=-\frac{(\omega
_{z}+\omega)}{\omega_{xy}}f_{z}$ and $f_{y}=-(\omega_{xy}\tau)^{-1}f_{z}$.
\ Note that, although we will not discuss these components further in this
letter, they are an integral part of the model and crucial for the
determination of $f_{z}$. \ An expression similar to Eqn. \ref{fzU} has been
used to describe Gilbert damping in a ferromagnet \cite{tserkovnyak2}, and
Eqn. \ref{fzU} can be viewed as a generalization of that result. \ In
resonance when $\omega_{z}=-\omega$, the universal value $\hbar\omega$ is
achieved when $\omega_{xy}\tau>>1$. \ Hereafter, we will focus on the more
general case, where a finite $(\omega+\omega_{z})$ strongly damps the spin
accumulation \cite{resonance}. \ Then, in the limit of long spin relaxation
times such that $\tau^{-1}<<(\omega_{xy},\omega_{z}+\omega)$, the prefactor
reduces to $(1+(\frac{\omega_{z}+\omega}{\omega_{xy}})^{2})^{-1}$ and is
independent of $\tau$; \ $f_{z}$ will then attain the universal value of
$\hbar\omega$ only when $\omega_{xy}>>(\omega_{z}+\omega)$.

We now turn to discussing the hybrid system of Fig. \ref{diagram}b. \ The
rotating field is only applied in the space $x<0$ (the "pumped" region I),
whereas for $x>0$ there are no magnetic fields (region II). \ We will consider
two physical realizations for the pumped region I: spin pumping in a
non-magnetic material with an applied, rotating magnetic field of magnitude
$B_{xy}=10\operatorname{mT}$, and $B_{z}=0$ (both region I and II are non-magnetic conductors); and a
model of a weak ferromagnet with $B_{xy}=1\operatorname{T}$ and $B_{z}=100\operatorname{T}$ representing the exchange interaction corresponding to small angle
magnetization precession around an easy axis (region II is a non-magnetic
conductor). This exchange interaction approach, in which the magnetization of
the localized d-moments couples to the s-electrons and polarizes them, has
been used in early spin-pumping theory \cite{silsbee} (which was applied to
transmission electron spin resonance), and more recently in the
mean-field-theory description of Gilbert damping \cite{tserkovnyak2}.

The system is solved as a boundary-value problem governed by Eqn.
\ref{BlochRot} with the diffusion term reinstated:%

\begin{equation}
\lambda^{2}\nabla^{2}\vec{f}=\vec{f}+(\vec{\omega}_{B}+\vec{\omega})\tau
\times\vec{f}+\hbar\vec{\omega}\times\vec{\omega}_{B}\tau.\label{Bloch3}%
\end{equation}
This equation can either be solved numerically directly, for instance using a
finite element method solver, or by using the boundary conditions to determine
the six constants of integration (for each region) of the general solution
\cite{ourselves}. \ For simplicity in our model system we have assumed that
the spin diffusion lengths, times and diffusion constants are the same for
both regions. \ With these assumptions the boundary conditions at the
interface reduce to continuity of $\vec{f}$ and $\nabla\vec{f}$ at $x=0$.
\ When the regions are unbounded, the outer boundary conditions are that in
region I, the bulk result (Eqn. \ref{fzU} and below) is recovered; in region
II, $\vec{f}(x>>0)=0$. \ When the regions are bounded, the boundary condition
is that there is no leakage of spin current, so $\nabla\vec{f}$ $=0$ at the
boundaries. \ In the calculations the diffusion constant has been fixed at
$D=0.002\operatorname{m}^{2}/\operatorname{s}$ (such as for Al) so the spin diffusion length $\lambda$ scales as
$\sqrt{\tau}$, and $\omega=10\operatorname{GHz}$.

\begin{figure}[ptb]
\begin{center}
\includegraphics[height=8.5339cm,width=6.5921cm]{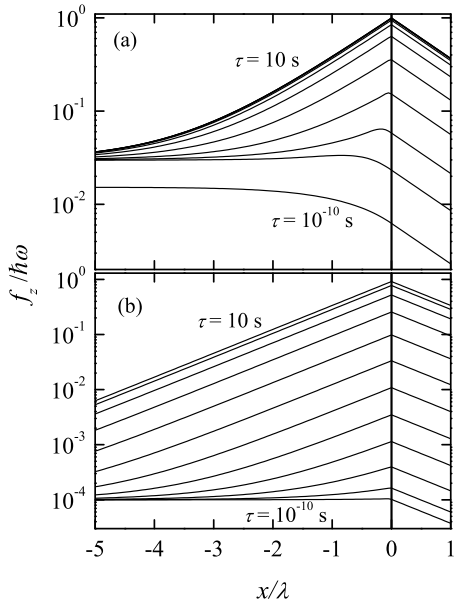}%
\caption{The spin accumulation $\ f_{z}/\hbar\omega$ as a function of
$x/\lambda$ away from the interface, for $\tau$ ranging in decades from 100 ps
to 10 s. \ Both regions are unbounded. \ For (a), $B_{xy}=10\operatorname{mT}$
and $B_{z}=0$, for (b) $B_{xy}=1\ $T and $B_{z}=100$ T.}%
\label{unbounded}%
\end{center}
\end{figure}

\begin{figure}[ptb]
\begin{center}
\includegraphics[height=5.2719cm,width=6.5921cm]{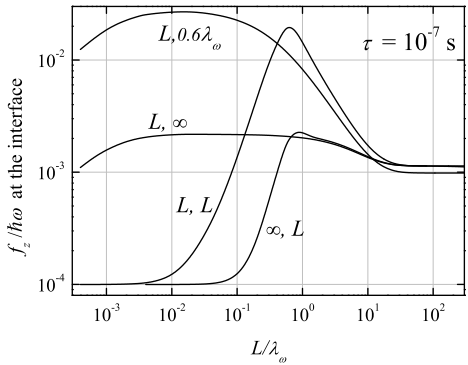}%
\caption{Spin accumulation at the interface vs the normalized bounding length
$L/\lambda_{\omega}$, calculated with $\tau=10^{-7}$ s under various bounding
conditions indicated as $L_{1},L_{2}$ in the figure. \ The largest enhancement
is obtained for the $L,0.6\lambda_{\omega}$ case with $L\simeq0.6\lambda
_{B}=0.6\sqrt{2\pi D/\omega_{B}}=15\operatorname{nm}$.}%
\label{summaryfzL}%
\end{center}
\end{figure}

\begin{figure}[ptb]
\begin{center}
\includegraphics[height=5.2983cm,width=6.5921cm]{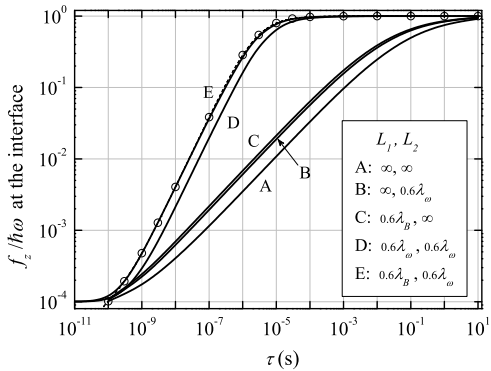}%
\caption{The spin accumulation at the interface vs $\tau$ for specific
bounding lengths indicated as $L_{1},L_{2}$ (curves A through\ E). The open
circles are based on Ref. \cite{brataas} with $\eta=10^{-5}\frac{1}{\sqrt
{\tau}}\tanh\sqrt{\frac{2\pi}{\omega\tau}}$ (see text for details).}%
\label{summaryfztsf}%
\end{center}
\end{figure}

We first consider the case where both regions are unbounded. \ \ Fig.
\ref{unbounded} shows the results of the calculation for the two different
field pumping regimes, plotting the spatial dependence of $f_{z}/\hbar\omega$
near the interface for values of $\tau$ ranging over many decades. \ The most
striking feature in Fig. \ref{unbounded} is that the spin accumulation at the
interface is strongly enhanced over the bulk, uniform system value (at
$x<<0$), and in the limit of very large $\tau$ achieves the universal value
$\hbar\omega,$ similar to the results of Ref. \cite{brataas}. \ The
calculations seem to describe a spin current injected at the interface which
diffuses into both regions, even though the pumping occurs uniformly
throughout region I.%
\begin{figure}[ptb]
\begin{center}
\includegraphics[height=5.2104cm,width=6.5921cm]{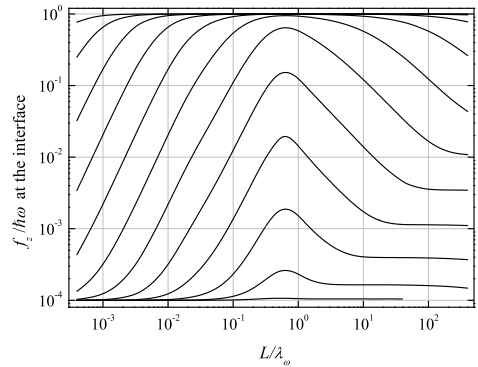}%
\caption{Spin accumulation at the interface for the symmetrically bounded
system plotted as a function of the normalized bounding length $L/\lambda
_{\omega}$, where $\lambda_{\omega}=\sqrt{2\pi D/\omega}\simeq
1.1\operatorname{\mu m}$, for $\tau$ ranging in decades from $10^{-10}$
(bottom curve) to 10 s (top curve).}%
\label{symmbounded}%
\end{center}
\end{figure}
The enhancement of the spin accumulation at the interface is the primary
result of this letter. \ In the unbounded system this effect is universal only
for unphysically large $\tau$, so we now try bounding the system in order to
optimize the effect. \ In what follows we will focus only on the weak
ferromagnet model, however most of the results apply generally to both
systems. \ Fig. \ref{symmbounded} shows the spin accumulation at the interface
obtained for symmetric bounding of the regions at $x=\pm L$ as a function of
$L/\lambda_{\omega}$, over decades of $\tau$ from $10^{-10}$ to $10$ s.\ The
spin accumulation is optimized when the sample dimension $L\simeq
0.6\lambda_{\omega}$, independent of $\tau$. \ Fig. \ref{summaryfzL} shows the
dependence of the spin accumulation on the bounding length, calculated under
various bounding conditions for a particular $\tau=10^{-7}$ s. \ Bounding only
region II yields optimal spin accumulation for $L/\lambda_{\omega}$ close to
one, however upon bounding only the pumped region the spin accumulation peaks
at a smaller length scale, corresponding to $L\simeq\lambda_{B}$. \ Bounding
region II at $L_{2}=0.6\lambda_{\omega}$ and allowing $L_{1}$ to vary yields a
peak at $L_{1}\simeq0.6\lambda_{B}$. \ The spin accumulation obtained for
different specific bounding lengths as a function of $\tau$ is summarized in
Fig. \ref{summaryfztsf}. \ The main message from this figure is that by
engineering the sample dimensions it is possible to achieve universality at
much lower $\tau$.


A qualitative explanation for the interface enhancement is the averaging
effect region II has on the x- and y-components of the spin accumulation.
\ These components diffuse into region II\ but can diffuse back after some
delay time. \ The back-flow will have a distribution of spin angles with
respect to the rotating field in region I that tends to average out their
influence, an effect similar to increasing the spin relaxation rate $\tau
_{xy}^{-1}$ for the x- and y-components of the spin accumulation. \ However
$f_{z}$ does not experience such an averaging effect. \ In a simple model of
spin accumulation in a uniform system it can be shown that reducing $\tau
_{xy}$ while staying in the limit $\omega_{xy}^{2}\tau_{xy}\tau_{z}>>1$ leads
to an enhancement of the spin accumulation \cite{ourselves}.

In the interface scattering matrix model \cite{brataas}, the spin accumulation
close to the interface is given as $f_{z}=\hbar\omega\frac{\sin^{2}\theta
}{\sin^{2}\theta+\eta}$, where $\theta$ is the precession cone angle and
$\eta=(\tau_{i}/\tau)\tanh(L/\lambda)/(L/\lambda)$ is a reduction factor which
depends on the normal metal properties and the spin-injection rate $\tau_{i}$
which is inversely proportional to the mixing conductance $g_{\uparrow
\downarrow}$. \ In the limit of very large $\tau$ and $\lambda$, the reduction
factor goes to zero and the cone angle dependence factors out, leaving the
maximum spin accumulation $f_{z}=\hbar\omega$. \ The authors restrict their
analysis to the case where the ferromagnet thickness should be less than the
spin diffusion length. \ To compare to their result we take $\theta\simeq
\frac{B_{xy}}{B_{z}}=\frac{1}{100}$ and evaluate $\eta$ for $L=\lambda
_{\omega}$, and find that $\eta$ depends on $\tau$ as $\eta\propto\frac
{1}{\sqrt{\tau}}\tanh\sqrt{\frac{2\pi}{\omega\tau}}$. \ This functional
dependence on $\tau$ agrees well with curves D and E, as shown for the fit to
curve E (with $10^{-5}$ as the fitting parameter for $\eta$), even though our
model is derived for an entirely different regime.

Our model of spin pumping in a non-magnetic region by a rotating magnetic
field can be applied quite generally to diffusive systems, such as metals and
semiconductors. \ \ Considering a semiconductor with $\tau$ as large as 100 ns
and experimentally quite accessible $\omega=10^{12}\operatorname{Hz}$, 
then a spin accumulation close to the universal result of $\hbar\omega$
corresponding to 700 $\operatorname{\mu V}$ can be achieved even in the non-resonant case. 
The measurement of small
spin accumulation signals due to the spin Hall effect in semiconductors by
Kerr rotation microscopy has recently been demonstrated \cite{kato}. \ The
spin pumping-induced spin accumulation and its interface enhancement that we
have described should be measurable in similar systems.

We acknowledge useful discussions with G. E. W. Bauer and M. Costache.
\ Support has been provided by the Stichting Funtamenteel Onderzoek der
Materie (FOM) \ and the RTN Spintronics Network.


\begin{thebibliography}{99}
\bibitem {magnetization}A spin accumulation $\vec{f}(x,t)$ is in general not
directly proportional to the magnetization. However, $f_{z}$ in Eqn.
\ref{fzU}, as a steady-state quantity, is associated with a contribution to
the magnetization of $\frac{1}{2}\mu_{B}N(E_{F})f_{z}$, where $N(E_{F})$ is
the density of states at the Fermi level.
\bibitem {quantum}Quantum-mechanically this cannot be fully justified, however
we work in a strictly classical ensemble limit such that temporal and spatial
coherence and interference effects can be neglected.
\bibitem {resonance}It is difficult to realize the resonant case
experimentally when pumping from a ferromagnet. \ The resonant case is treated
elsewhere for pumping from a non-magnetic material \cite{ourselves}.
\end{thebibliography}
\end{document}